\documentclass{PoS}
\usepackage[utf8]{inputenc}
\usepackage{amsfonts}
\usepackage{amsmath}
\usepackage{amssymb}
\usepackage[pdftex]{graphicx}

\title{A study of the radiative transition $\pi \pi \to \pi \gamma^{\ast}$ with lattice QCD}

\ShortTitle{A study of the radiative transition $\pi \pi \to \pi \gamma^{\ast}$ with lattice QCD}

\author{\speaker{Luka Leskovec}$^{a}$,
                Constantia Alexandrou$^{b,c}$,
                Giannis Koutsou$^{c}$, 
                Stefan Meinel$^{a,d}$, 
                John W. Negele$^{e}$, 
                Srijit Paul$^{c,f}$, 
                Marcus Petschlies$^{g}$, 
                Andrew Pochinsky$^{e}$, 
                Gumaro Rendon$^{a}$, 
                Sergey Syritsyn$^{d,h}$\\
      ${^a}$Department of Physics, University of Arizona, Tucson, AZ 85721, USA\\
      ${^b}$Department of Physics, University of Cyprus, CY 1678 Nicosia, Cyprus\\
      ${^c}$Computation-based Science and Technology Research Center, The Cyprus Institute, CY 2121 Nicosia, Cyprus\\
      ${^d}$RIKEN BNL Research Center, Brookhaven National Laboratory, Upton, NY 11973, USA\\
      ${^e}$Center for Theoretical Physics, Massachusetts Institute of Technology, Cambridge, MA 02139, USA\\
      ${^f}$Department of Mathematics and Natural Sciences, University of Wuppertal, D-42119 Wuppertal, Germany\\
      ${^g}$Helmholtz-Institut f\"ur Strahlen- und Kernphysik, University of Bonn, D-53115 Bonn, Germany\\
      ${^h}$Department of Physics and Astronomy, Stony Brook University, Stony Brook, NY 11794, USA\\
      E-mail: \email{leskovec@email.arizona.edu}}

\abstract{     
Lattice QCD calculations of radiative transitions between
hadrons have in the past been limited to processes of hadrons stable under the
strong interaction. Recently developed methods for $1\to2$ transition matrix
elements in a finite volume now enable the determination of radiative decay
rates of strongly unstable particles. Our lattice QCD study focuses on the
process $\pi \pi \to \pi \gamma^{\ast}$, where the $\rho$ meson is present as
an enhancement in the cross-section. We use $2+1$ flavors of clover fermions at a pion mass of approximately $320$ MeV and a lattice size of
approximately $3.6$ fm. The required $2$-point and $3$-point correlation
functions are constructed from a set of forward, sequential and stochastic
light quark propagators. In addition to determining the $\rho$ meson resonance
parameters via the Lüscher method, the scattering phase shift is used in
conjunction with the $1\to2$ transition matrix element formalism of Brice\~no,
Hansen and Walker-Loud \cite{Briceno:2014uqa} to compute the
$\pi\pi\to\pi\gamma^{\ast}$ amplitude at several values of the momentum
transfer and $\pi\pi$ invariant mass. 
}

\FullConference{34th annual International Symposium on Lattice Field Theory\\
        24-30 July 2016\\
        University of Southampton, UK}

\begin{document}

\section{Introduction} 

Lattice QCD studies in the past have focused on calculations of hadronic
masses within the single-hadron approach, resonance masses and their
corresponding strong decay widths via the L\"uscher method
\cite{Luscher:1990ux}, as well as form factors and matrix elements involving
transitions between hadrons stable under the strong interaction. There were
also some exploratory calculations of matrix elements involving unstable
hadrons, where the effects of the strong decay were neglected
\cite{Crisafulli:1991pn,Horgan:2013pva}, leading to uncontrolled finite volume
effects.

The effect of the multi-hadron state for the $K\to \pi\pi$ decay was first
described by  Lellouch and L\"uscher \cite{Lellouch:2000pv} and was extended
to all elastic states below the inelastic threshold \cite{Lin:2001ek}. The
Lellouch-L\"uscher factor, which encodes the finite volume effects that affect
the transition matrix elements, has been generalized to describe also hadrons
in a moving frame \cite{Christ:2005gi}. The inclusion of multiple decay
channel modes of the unstable hadron have been addressed in
\cite{Hansen:2012tf} and a specific setup to calculate the $\Delta \to N
\gamma$ decay was proposed in \cite{Agadjanov:2014kha}.  A recent paper by
Brice\~no, Hansen and Walker-Loud \cite{Briceno:2014uqa} was the first to
derive the effects in quantum field theory as well as put forward a full
general setup for $1\to2$ transition matrix elements involving an arbitrary
number of coupled two-hadron channels.  The first study employing the formalism of
Brice\~no, Hansen and Walker-Loud (BHWL) was performed by the Hadron Spectrum
collaboration only recently \cite{Briceno:2015dca, Briceno:2016kkp}, for the
$\pi \pi (\to \rho) \to \pi \gamma$ transition amplitude. In our work we study
the same channel, but use different methods to construct the correlation
functions.

\section{On the Brice\~no-Hansen-Walker-Loud formalism} 

The finite volume effects of the transition matrix element we consider in this work are described by
\begin{align} 
\label{LL} 
\frac{|F_{IV}( \nu; J_{\mu}^{QED}; q^2, s_{\pi\pi})|^2}{|\langle n, \Lambda, \nu, \vec{p}_{\pi\pi} | J_{\mu}^{QED} | \pi, \vec{p}_{\pi} \rangle|^2} &= \frac{32\pi E_{\pi} \sqrt{s_{\pi\pi}}}{k} \bigl[ \frac{\partial \delta_1(\sqrt{s_{\pi\pi}})}{\partial E_{\pi\pi}} + \frac{\partial \phi_1^{\vec{p}_{\pi\pi}}(k)}{\partial E_{\pi\pi}} \bigr]
\end{align} 
where $\phi_1^{\vec{p}_{\pi\pi}}(k)$ comes from the quantization condition of
the L\"uscher method, $\cot{\delta_1} + \cot{\phi_1^{\vec{p}_{\pi\pi}}(k)}=0$,
$k$ is related to $\sqrt{s_{\pi\pi}}$ via $\sqrt{s_{\pi\pi}}=2\sqrt{m_{\pi}^2
+ k^2}$, $\delta_1$ is the phase shift that describes the $\pi\pi$ resonant
scattering, $\langle n, \Lambda, \nu, \vec{p}_{\pi\pi} | J_{\mu}^{QED} | \pi,
\vec{p}_{\pi} \rangle$ is the finite volume matrix element and $F_{IV}$ is the
infinite volume matrix element, which can be further reduced by taking into
account the Lorentz symmetry. $E_{\pi\pi}$ is the energy of the
given state $n$ in the irrep $\Lambda$ as determined by the spectrum study.

The factor on the right-hand-side is referred to as the Lellouch-L\"uscher
factor \cite{Lellouch:2000pv} and represents the mapping from the finite
volume matrix element to the infinite volume transition amplitude. The basic
idea of the BHWL formalism is to  calculate the Lellouch-L\"uscher factor
using prior knowledge  and then perform the mapping from the finite volume to
the infinite volume. This is done not only for the ground state, but
for all relevant states in the lattice irrep in which the unstable
hadron, in our case the $\rho$, is present.

Thus, an essential part of the BHWL formalism is the spectroscopy study,  where
the phase shift in a given channel (in our case elastic $\pi\pi$ scattering)
is calculated with the L\"uscher method and its generalizations
\cite{Luscher:1990ux}. This provides us with two quantities - the phase shift
needed to determine the mapping, and the specific linear combinations of
interpolating operators that create a well defined specific state. These
combinations yield the optimized correlators \cite{Dudek:2009kk}, and
give us access to the matrix elements at several different values of
the  $\pi\pi$ invariant mass. A simple outline of the procedure we employ to
determine the infinite volume amplitude looks like:

\begin{itemize}     
    \item[1)] perform a lattice calculation of $\pi\pi$ $p$-wave resonant
scattering in the multi-hadron approach and determine the phase shift with the
L\"uscher method     
    \item[2)] determine the parameters of the Breit-Wigner phase shift form and
calculate the derivative of the phase shift with respect to the two-particle energy for the given Breit-Wigner parametrization
    \item[3)] calculate the $\pi\pi \to \pi \gamma$ three point functions
involving both $\bar{q}\Gamma_i q$ and $\pi\pi$ interpolating operators and utilize the
information from 1) to project them to definite states
    \item[4)] use the derivative of the phase shift from 2) combined with the function
$\frac{\partial \phi_1^{\vec{p}_{\pi\pi}}(k)}{\partial E_{\pi\pi}}$ to determine the mapping from the  finite volume matrix elements to the
infinite volume amplitude.

\end{itemize}

\section{Construction of the correlation functions} 
\label{sec:method}


To calculate the $2$-point and $3$-point correlation functions we adapted a
method using a combination of forward, sequential, and stochastic propagators.
The forward quark propagator $S_f$ from the initial point $(\vec{x}_i, t_i)$
to the final point $ (\vec{x}_f, t_f)$ is the inverse of the Dirac operator
$D$:

\vspace{-0.1cm}
\begin{minipage}[tb]{0.21\textwidth}
  \begin{center}
    \vspace{+0.5cm}
    \includegraphics[width=\textwidth]{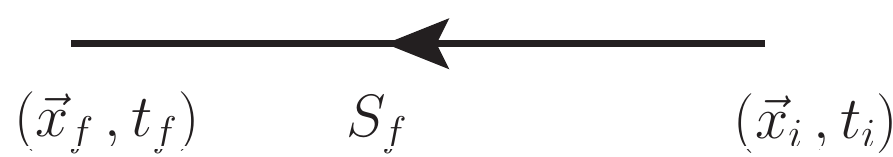}
  \end{center}
\end{minipage}
\begin{minipage}[tb]{0.73\textwidth}
  \begin{center}
    \begin{equation}
      S_f(\vec{x}_f,t_f; \vec{x}_i,t_i)_{\alpha\beta}^{ab} = D^{-1}(\vec{x},t_f;\vec{x}_i,t_i)_{\alpha\beta}^{ab}\,,
    \end{equation} 
  \end{center}
\end{minipage}
\vspace{0.3cm}

\noindent where $\alpha,\beta$ are spin indices and $a,b$ are color indices. The
sequential propagator describes a quark flow through a vertex of a given flavor
and Lorentz structure. It is obtained by using the product of a forward propagator and the gamma matrix in the vertex as a the source for the solver. It requires
$12$ inversions (one for each spin and color index) for each distinct
sequential source time $t_{seq}$, momentum $\vec{p}$ and and $\Gamma$ matrix:

\begin{minipage}[tb]{0.21\textwidth}
  \begin{center}
    \vspace{+0.5cm}
    \includegraphics[width=\textwidth]{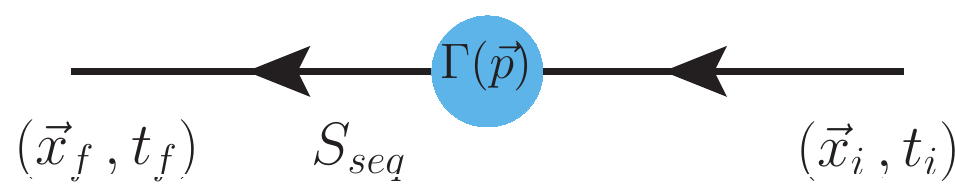}
  \end{center}
\end{minipage}
\begin{minipage}[tb]{0.73\textwidth}
  \begin{center}
    \begin{align}
      & S_{seq}(\vec{x}_f,t_f;\vec{x}_i,t_i; t_{seq},\vec{p},\Gamma) \nonumber\\
      & \quad = \sum\limits_{\vec{x}_{seq}}\,D^{-1}(\vec{x}_f,t_f; \vec{x}_{seq},t_{seq})\,\Gamma\,\mathrm{e}^{i\vec{p}\vec{x}_{seq}}\,S_f(\vec{x}_{seq},t_{seq};\vec{x}_i,t_i)\,.
      \label{eq:sequential_propagator}
    \end{align}
  \end{center}
\end{minipage}

\noindent The stochastic timeslice-to-all propagator is defined as by the inversion of
the Dirac matrix on a stochastic timeslice momentum source:

\begin{minipage}[tb]{0.21\textwidth}
  \begin{center}
    \vspace{+0.5cm}
    \includegraphics[width=\textwidth]{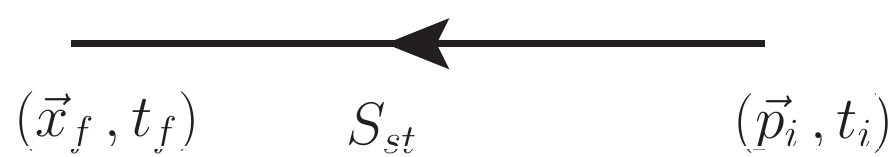}
  \end{center}
\end{minipage}
\begin{minipage}[tb]{0.73\textwidth}
  \begin{center}
    \begin{align}
      \phi(t_i, \vec{p}) &= D^{-1}\,\xi(t_i,\vec{p}) 
      \label{eq:S_st}\\
      \xi(t_i,\vec{p}_i)_{t,\vec{x},\beta,b} &= \sum\limits_{\vec{y}} \delta_{t,t_i}\,\mathrm{e}^{i\vec{p}_i\vec{y}}\,\xi(t_i)_{\vec{y},\beta,b}\,. \nonumber
      \end{align}
  \end{center}
\end{minipage}

\noindent This technique provides an efficient way to evaluate the partially
disconnected diagrams involved in typical multi-hadron studies with reasonable
cost. In addition to time-dilution of the stochastic momentum source, we also
apply spin-dilution to make use of the efficient one-end-trick in our
contractions:
\begin{align}
  \xi(t_i,\vec{p}_i,\alpha)_{t,\vec{x},\beta,b} &= \sum\limits_{\vec{y}} \delta_{t,t_i}\,\delta_{\alpha,\beta}\,
    \mathrm{e}^{i\vec{p}_i\vec{y}}\,\xi(t_i)_{\vec{y},b}\,.
    \label{eq:S_st_oet}
\end{align}
To enhance the overlap to the low lying states contributing to a correlator we apply
source and sink smearing to the propagator types listed above by replacing $
S_X \to W\,S_X\,W^\dagger $, where $W$ denotes the Wuppertal-smearing operator
\cite{Gusken:1989ad} and $X \,\in\,\left\{ f,\,seq,\,st \right\}$.


To determine the $\pi\pi$ $p$-wave phase shift we calculate the $2$-point
correlation functions in the multi-hadron approach, within which we build a
correlation matrix $C_{ij}(t)=\langle O_i(t) O_j^{\dagger}(0)\rangle$ from two
types of interpolating operators $O^{\bar{q}q}$ and $O^{\pi\pi}$:
\begin{align}
O_{\nu}^{\bar{q}q} &\sim \bar{q}\Gamma_\nu q (\vec{p}_{\pi\pi}), \cr
O_{\nu}^{\pi\pi} &\sim \pi^+(\vec{p}_{1})\pi^-(\vec{p}_{2}),
\end{align}
where $\vec{p}_{\pi\pi}=\vec{p}_{1}+\vec{p}_{2}$ and $\Gamma_\nu = \{ \gamma_\nu,
\gamma_4 \gamma_\nu \}$. We project these correlation functions to definite $\pi\pi$
momenta, where $\vec{p}_{\pi\pi} = \frac{2\pi}{L}(0,0,0)$,
$\frac{2\pi}{L}(0,0,1)$, $\frac{2\pi}{L}(0,1,1)$ and
their permutations. We combine the interpolators in such a manner that each
is in a well defined irreducible representation, which are constructed by the
projection operator defined in Eq. (13) of \cite{Feng:2010es}. 



\begin{figure}[!htb]
  \centering
  \includegraphics[width=0.71\textwidth]{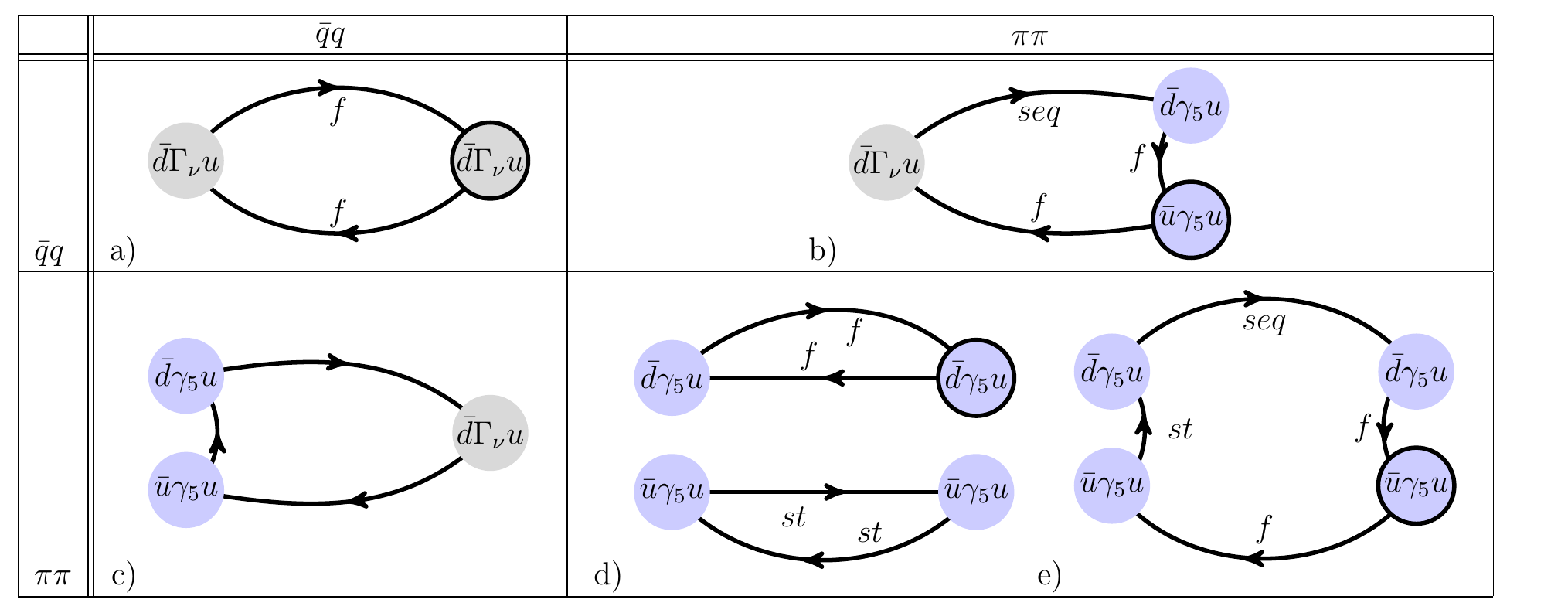}
  \caption{\label{fig:wick} The correlation matrix $C_{ij}$ for the $2$-point
functions, which involves $4$ types of diagrams that are built from the
propagators defined in Section \protect\ref{sec:method}.}
\end{figure}

Schematically, the Wick diagrams involved in constructing the correlation
matrix  are presented in Figure \ref{fig:wick} and are built using the
propagators defined above.

The spectrum in each of the momentum frames and irreps is obtained by
solving the generalized eigenvalue problem (for details on GEVP
see \cite{Blossier:2009kd} and \cite{Orginos:2015tha}) $C(t)u_n(t) =
\lambda_n(t,t_0)C(t_0)u_n(t)$. We use $t_0=2$ and checked that for $t_0$ up to
$6$ the spectrum is consistent with our choice.

The  $(I)J^{PC}=(1)1^{--}$ elastic $\pi\pi$ scattering phase shift is obtained
using the L\"uscher method \cite{Luscher:1990ux} as well as its
generalizations \cite{Rummukainen:1995vs,Briceno:2014oea}. The relevant
equations for the mapping from the finite volume to the infinite volume are
listed in \cite{Dudek:2012xn}, and a comparison of previous studies can be found in
\cite{Bali:2015gji}.

\begin{figure}[!htb]
  \centering
  \includegraphics[width=0.71\textwidth]{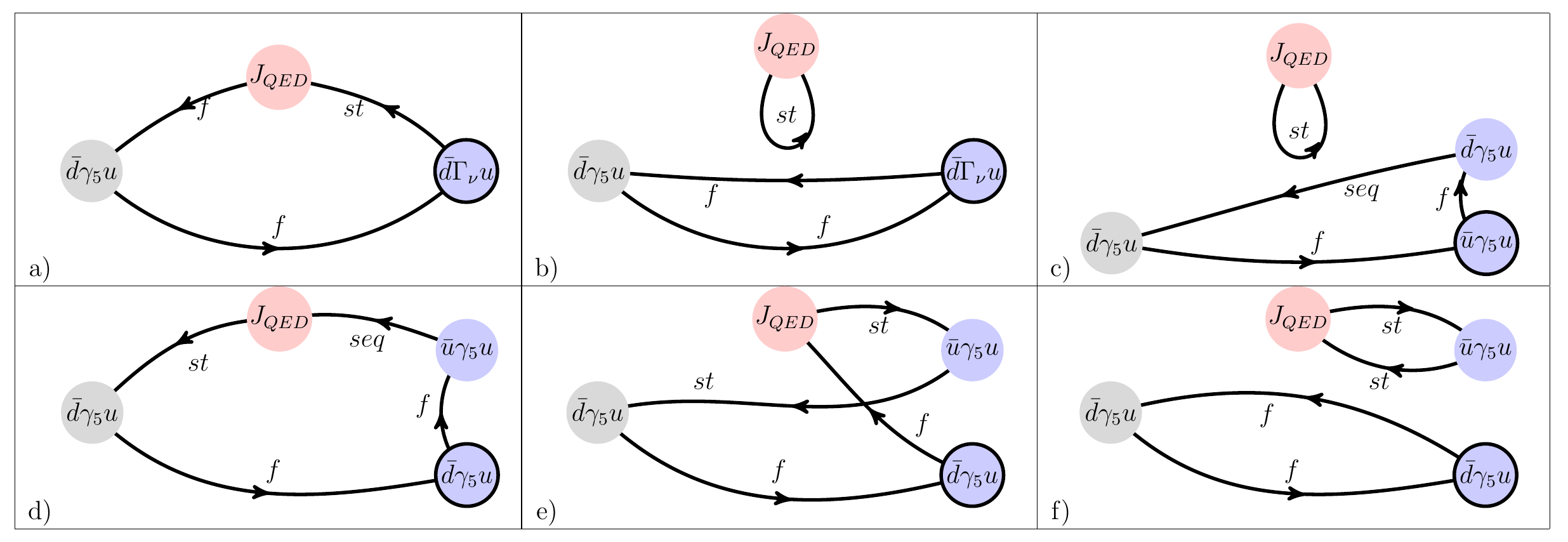}
  \caption{\label{fig:wick3} Wick diagrams for the $3$-point correlation functions. We have not yet included the disconnected diagrams in b) and c) to the correlation function.}
\end{figure}

To determine the $3$-point correlation functions $C_{3}$ that project to a
definite state in the irreducible representation, we calculate the $3$-point
functions $C_3^i$:
\begin{align}
C^i_3(t_J,t_{i},t_{f})
=\sum_{n \in \Lambda,\vec{p}_{\pi\pi}} \langle 0|  O_{i} |n,\Lambda,\nu \rangle \langle n, \Lambda, \nu, | J^{\mu}_{QED} | \pi,\vec{p}_{\pi} \rangle \langle \pi |O_{\pi} | 0 \rangle 
&\frac{e^{-E_n(t_{f} - t_J)} e^{-E_{\pi}(t_J - t_{i})}}{2E_{\pi} E_n},
\end{align}

\noindent where $t_J$ is the current insertion timeslice, $t_{i}$ is the pion creation
timeslice and $t_{f}$ is the $\pi\pi$ channel annihilation timeslice.
$J_{\mu}^{QED}$ is the QED current, $J_{\mu}^{QED} = Z_V(\frac{2}{3}
\bar{u}\gamma_\mu u - \frac{1}{3}\bar{d}\gamma_\mu d)$ and $Z_V= 0.79700(24)$.
The index $i$ runs over the interpolators used in the $2$-point function part
and includes both the $\pi\pi$ and $\bar{q}q$ interpolators to describe the
$\rho$/$\pi\pi$ state.  All the relevant $3$-point functions are shown in Fig. \ref{fig:wick3}, however we have not yet included the disconnected terms in Fig. \ref{fig:wick3}b) and c).

The projection to the definite state is done by using the idea of the
optimized correlator
\cite{Dudek:2009kk,Becirevic:2014rda,Shultz:2015pfa} to construct combination
of the $C_3^i$ that will have the best possible overlap to a given state,
$C_3(n,t_J,t_{i},t_{f}) = u_n^{i} C_3^i(t_J,t_{i},t_{f})$. To determine the matrix elements $\langle n,
\Lambda, \nu, \vec{p}_{\pi\pi} | J_{\mu}^{QED}| \pi, \vec{p}_{\pi} \rangle$,
where $n$ marks the state in the irrep $\Lambda$ with polarization $\nu$ and
momentum $\vec{p}_{\pi\pi}$, from the $3$-point functions, we use a ratio
similar to those defined in \cite{Detmold:2015aaa}:

\begin{align}
\nonumber
&R(n,t_J,t_{i},t_{f}) = \\
&\frac{C_3(n,t_J,t_{i},t_{f}) C_3^*(n,t_{f} + t_{i} - t_J,t_{i},t_{f})}{C_2^{(n)}(t_{f}, t_{i}) C_2^{(\pi)}(t_{f},t_{i}) } \; \to \; |\langle n, \Lambda, \nu,\vec{p}_{\pi\pi}| J^{\mu}_{QED} | \pi,\vec{p}_{\pi} \rangle|^2.
\end{align}

We determine the amplitude $|F_{IV}( \nu; J_{\mu}^{QED}; q^2, s_{\pi\pi})|$ for the $\pi\pi \to \pi \gamma$ transition by mapping
the finite volume matrix elements to the infinite volume matrix elements using
Eq. (\ref{LL}) and then performing the Lorentz invariant decomposition:
\begin{align}
F_{IV}( \nu; J_{\mu}^{QED}; q^2, s_{\pi\pi}) =
f_{\pi\pi,\pi}(q^2,s_{\pi\pi})\epsilon_{\mu\nu\alpha\beta}(p_{\pi\pi})_\alpha 
(p_{\pi})_\beta,
\end{align}
where $q=p_{\pi} - p_{\pi\pi}$ and $f_{\pi\pi,\pi}$ is the Lorentz invariant amplitude.

\section{Preliminary results and conclusions}

Our results are obtained on a $32^3 \times 96$ lattice gauge ensemble with
$N_f=2+1$ dynamical clover-Wilson fermions, which is described in detail in
ref. \cite{Green:2015wqa}. The light quark mass corresponds to a pion mass of
$317(2)$ MeV with a lattice spacing of $a=0.11403(77)$ fm. The presented results are
obtained from a subset of $N_{conf}=367$ configurations available on this lattice.

\begin{figure}[!htb]
  \centering
  \includegraphics[width=0.63\textwidth]{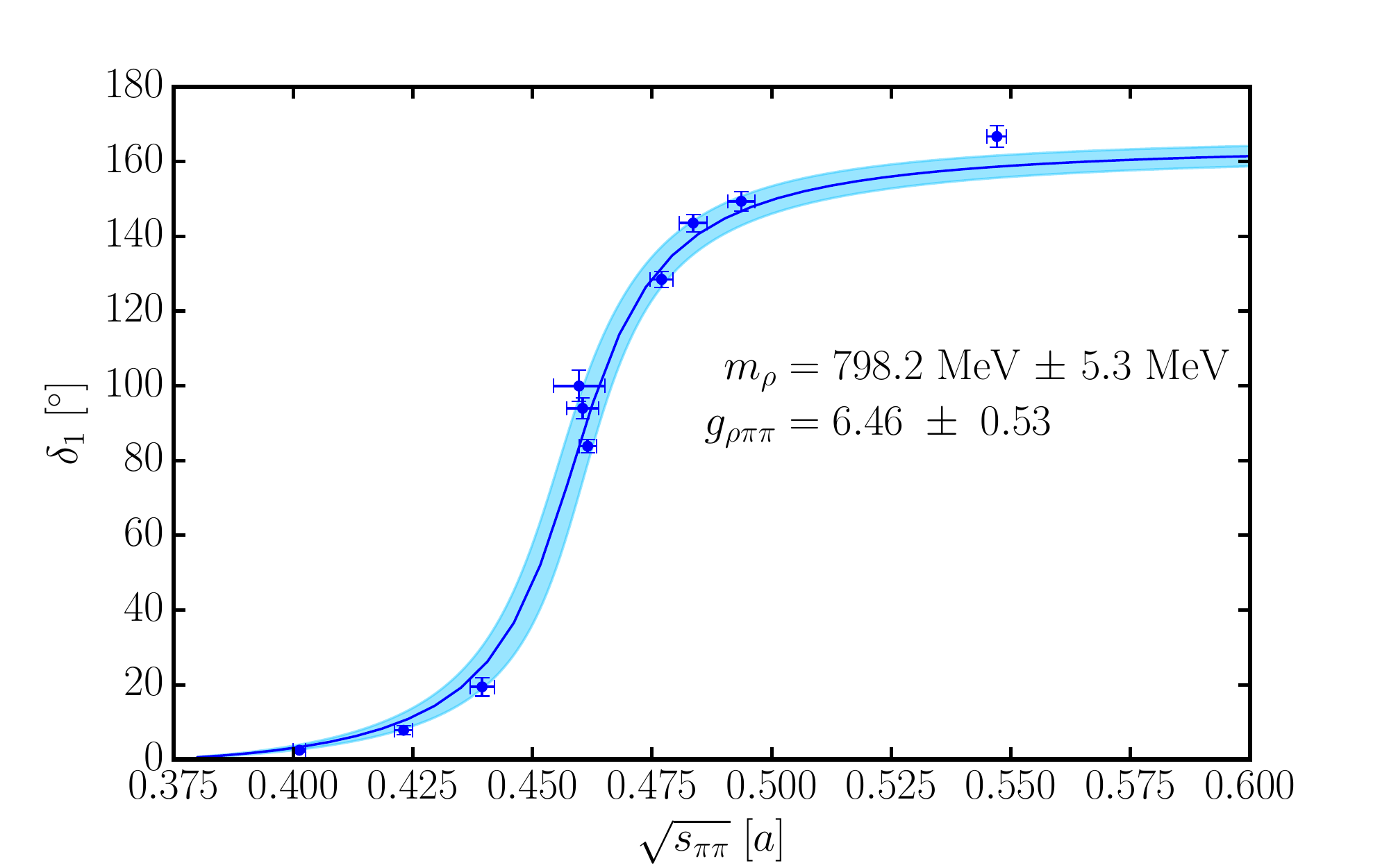}
  \caption{\label{fig:BW} The $\pi\pi$ scattering phase shift as calculated on our gauge ensemble with $m_{\pi}=317(2)$ MeV.}
\end{figure}

The $(I)J^{PC}=(1)1^{--}$ elastic $\pi\pi$ scattering phase shift is shown
in Fig. \ref{fig:BW}. By fitting a Breit-Wigner form to the phase shift we obtain
the following parameters for the $\rho$ resonance, which is present in this
channel:

\begin{align}
\tan \delta_1(\sqrt{s_{\pi\pi}}) = \frac{\sqrt{s_{\pi\pi}} \; \Gamma(\sqrt{s_{\pi\pi}})}{m_\rho^2 - s_{\pi\pi}}, \quad \quad \Gamma(\sqrt{s_{\pi\pi}})= \frac{g_{\rho\pi\pi}^2}{6\pi} \frac{k^3}{s_{\pi\pi}} \\
m_\rho = 798.2(5.3) \text{ MeV} \quad \quad g_{\rho\pi\pi}=6.46(53)
\end{align}
\vspace{-0.8cm}

\begin{figure}[!htb]
  \centering
  \includegraphics[width=0.55\textwidth]{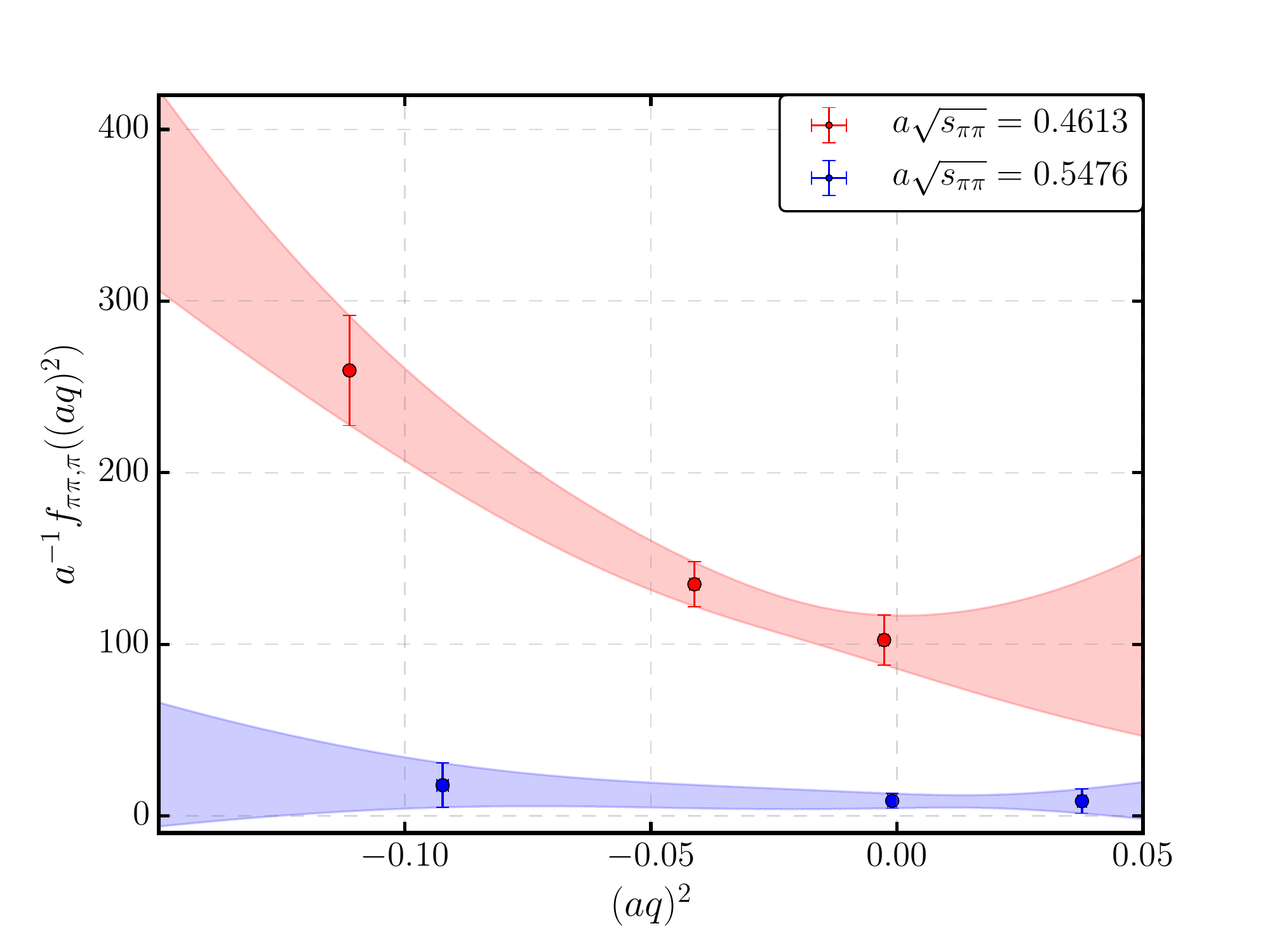}
  \caption{\label{fig:amp} The $\pi \pi \to \pi \gamma$ infinite volume amplitude as determined in the BHWL formalism using $t_{f} - t_{i}=10$. It exhibits an enhancement for $\sqrt{s_{\pi\pi}}$ in the vicinity of the $\rho$ resonance.}
\end{figure}

The $\pi \pi \to \pi \gamma$ infinite volume amplitude is shown in Fig.
\ref{fig:amp}. It exhibits the basic expected features, including an amplification near
$\sqrt{s_{\pi\pi}}\approx m_{\rho}$. Our results are comparable with the
results in a previous study by the Hadron Spectrum collaboration
\cite{Briceno:2015dca,Briceno:2016kkp}.
The adapted method for constructing correlation functions is computationally efficient on large volumes and produces good quality of data.

\section{Acknowledgement}
We are grateful to Kostas Orginos for providing the gauge ensemble generated with resources provided by XSEDE, which is supported by National Science Foundation grant number ACI-1053575. Our calculations were performed at NERSC, supported by the U.S. DOE under Contract No. DE-AC02-05CH11231. SM and GR are supported by NSF grant PHY-1520996. SM and SS also thank the RIKEN BNL Research Center for support. JN  was supported in part by the DOE Office of Nuclear Physics under grant \#DE-FG02-94ER40818. AP was supported in part by the U.S. Department of Energy Office of Nuclear Physics under grant \#DE-FC02-06ER41444. SP is supported by  the Horizon 2020 of the European Commission research and innovation programme under the Marie Sklodowska-Curie grant agreement No. 642069.


\end{document}